\documentclass[12pt]{article}
\usepackage{amssymb}
\usepackage{amsmath}
\usepackage{amscd}
\usepackage{latexsym}

\catcode`@=11
\renewcommand{\section}{\@startsection{section}{1}{0pt}{\medskipamount}
{\medskipamount}{\large\bf}}
\numberwithin{equation}{section}
\catcode`@=12

\def\be{\begin{equation}}
\def\ee{\end{equation}}
\def\Tr {{\rm Tr}}
\def \bear {\begin{eqnarray}}
\def \eear {\end{eqnarray}}

\def\Xint#1{\mathchoice
   {\XXint\displaystyle\textstyle{#1}}%
   {\XXint\textstyle\scriptstyle{#1}}%
   {\XXint\scriptstyle\scriptscriptstyle{#1}}%
   {\XXint\scriptscriptstyle\scriptscriptstyle{#1}}%
   \!\int}
\def\XXint#1#2#3{{\setbox0=\hbox{$#1{#2#3}{\int}$}
     \vcenter{\hbox{$#2#3$}}\kern-0.5\wd0}}

\newcommand{\sfrac}[2]{{\textstyle\frac{#1}{#2}}}
\newcommand{\half}{{\sfrac12}}

\renewcommand{\and}{{\quad{\rm and}\quad}}

\newcommand{\beq}{\begin{equation}}
\newcommand{\eeq}{\end{equation}}
\newcommand{\bea}{\begin{eqnarray}}
\newcommand{\eea}{\end{eqnarray}}

\advance \textheight by \topskip
\makeatletter
\@addtoreset{equation}{section}

\makeatother

\begin{document}
%%%%%%%%%%%%%%%%%%%%%%%%%%%%%%%%%%%%%%%%%%%%%%%
%\fontfamily{pnb}\fontsize{12pt}{16pt}\selectfont
%\fontfamily{pzc}\fontsize{14pt}{16pt}\selectfont
%\fontfamily{pbk}\fontsize{12pt}{16pt}\selectfont
\fontfamily{cmr}\fontsize{11pt}{14pt}\selectfont
%\fontfamily{phv}\fontshape{ro}\fontsize{11pt}{14pt}\selectfont
%\fontfamily{ptm}\fontseries{m}\fontshape{r}\fontsize{12pt}{16pt}\selectfont
%\fontfamily{pnc}\fontseries{m}\fontshape{r}\fontsize{11pt}{15pt}\selectfont
%\fontfamily{ppl}\fontseries{m}\fontshape{r}\fontsize{11pt}{15pt}\selectfont
%\usefont{T1}{phv}{m}{it}
%%%%%%%%%%%%%%%%%%%%%%%%%%%%%%%%%%%%%%%%%%%%%%%
\def \CMP {{Commun. Math. Phys.}}
\def \PRL {{Phys. Rev. Lett.}}
\def \PL {{Phys. Lett.}}
\def \NPBProc {{Nucl. Phys. B (Proc. Suppl.)}}
\def \NP {{Nucl. Phys.}}
\def \RMP {{Rev. Mod. Phys.}}
\def \JGP {{J. Geom. Phys.}}
\def \CQG {{Class. Quant. Grav.}}
\def \MPL {{Mod. Phys. Lett.}}
\def \IJMP {{ Int. J. Mod. Phys.}}
\def \JHEP {{JHEP}}
\def \PR {{Phys. Rev.}}
\def \JMP {{J. Math. Phys.}}
\def \GRG{{Gen. Rel. Grav.}}
%%%%%%%%%%%%%%%%%%%%%%%%%%%%%%%%%%%%%%%%%%%%%%%
%%%%%%%%%%%%%%%%%%%%%%%%%%%%%%%%%%%%%%%%%%%%%%%
\begin{titlepage}
\null\vspace{-62pt} \pagestyle{empty}
\begin{center}
\rightline{CCNY-HEP-13/4}
\rightline{June 2013}
\vspace{1truein} {\Large\bfseries
Effective action and phase transitions of scalar field on the fuzzy sphere}\\
\vspace{6pt}
\vskip .1in
{\Large \bfseries  ~}\\
\vskip .1in
{\Large\bfseries ~}\\
%%%%%%%%%%%%%%%%%%%%%%%%%%%%%%%%%%%%%%%%%%%%%%%%\vspace{.6in}
{\Large Alexios P. Polychronakos}\\
\vskip .2in
\noindent {\em
CCPP, Department of Physics, NYU \\
4 Washington Pl., New York, NY 10016, USA \\
\vskip 0.2cm
{\rm{and}} \\
\vskip 0.2cm
Physics Department, The City College of the CUNY \\
160 Convent Avenue, New York, NY 10031, USA\footnote{Permanent address}}\\
%{Email: alexios@sci.ccny.cuny.edu}

%{\itshape Physics Department\\
%City College of the CUNY\\
%New York, NY 10031}\\
\vskip .1in
{\fontfamily{cmtt}\fontsize{11pt}{15pt}\selectfont alexios@sci.ccny.cuny.edu}

\fontfamily{cmr}\fontsize{11pt}{15pt}\selectfont
\vspace{.8in}
%\vspace{1.5in}%\vspace{0.3in}
%%%%%%%%%%%%%%%%%%%%%%%%%%%%%%%%%%%%%%%%%%%%%%%%%%%%%%%%%%%%
\centerline{\large\bf Abstract}
\end{center}
Scalar field theory on the fuzzy two-sphere, represented as a
hermitian matrix model that includes kinetic, mass and quartic interaction terms, is studied. The effective action in the symmetric large-$N$ regime is analyzed using a self-consistent bootstrap method which fixes its form up to sixth order in the eigenvalues
and gives a closed expression to all orders in the quadratic invariant of the matrix, valid close to semicircular
distributions. Using this action the eigenvalue distribution is calculated for the interacting theory in the appropriate
scaling limit and the phase transition from the disordered to the symmetric ordered phase is identified, 
including nonperturbative effects.
\end{titlepage}
%%%%%%%%%%%%%%%%%%%%%%%%%%%%%%%%%%%%%%%%%%%%%%%%%%%%%%
\section{Introduction and conclusions}

Noncommutative field theory presents an interesting setting for calculating properties of quantum fields without the
burden of ultraviolet divergences. The noncommutative scale could either arise from fundamental physics
(such as quantum gravity) or simply be a convenient regularization method. Effects such as UV-IR mixing complicate somewhat
the picture but also manifest analogies with planar gauge theory. As a result, noncommutative spaces and field theories
have been an important topic of research for quite a while \cite{douglas}.
Such spaces arise as brane solutions in string theory and in the matrix version of $M$-theory \cite{taylor}.
Noncommutative gauge theories are especially interesting since they can describe fluctuations of the brane solutions 
and unify in a natural way gauge and spatial degrees of freedom.

Compact noncommutative spaces are particulary attractive, as they constitute both UV and IR
regularizations that preserve many of the symmetries of the commutative theory. Due to this double cutoff, field theory on
these spaces has finitely many degrees of freedom. The simplest such space is the noncommutative two-sphere,
also called ``fuzzy sphere" \cite{BerMad}. This is the lowest of a tower of (even dimensional) fuzzy spaces corresponding to
noncommutative versions of $CP^n$. On such spaces, fields become matrices and field theory effectively becomes
a matrix model.

Matrix models have been a subject of interest in physics and mathematics for a long time, and much
of the machinery developed for solving them can be imported in the study of noncommutative physics.
The first and best known results were derived by Wigner in his work on Gaussian matrix ensembles \cite{wigner}. 
Assuming a normal distribution for the elements of the matrix,
the classic semicircle eigenvalue distribution of Wigner emerges.
Matrix models have since emerged in many other contexts in physics: modeling Riemann manifolds in string theory \cite{string}, integrable systems such as the Calogero model \cite{CalPol}, quantum Hall states \cite{QH}, other condensed matter systems \cite{guhr}, quantum chaos \cite{chaos}, two-dimensional Yang-Mills theory \cite{steinYM} etc.

The matrix model corresponding to scalar field theory on the fuzzy sphere includes a term that represents the
noncommutative version of the Laplacian (kinetic term), in addition to mass and interaction terms. The kinetic
term is not invariant under unitary conjugations, unlike the remaining terms, and makes the model nontrivial.
A diagrammatic study of this theory on the noncommutative plane with a cutoff was done in \cite{cutoff}.
Numerical simulations of the model on the fuzzy sphere were performed in \cite{numer} and phase transitions in the
large-$N$ limit were identified.

The question of the eigenvalue distribution and phase structure of the scalar field model in the presence of
the kinetic term and interactions
can best be dealt with in terms of the effective action for the eigenvalues \cite{{steinE},{ocon},{Sae}}.
In \cite{ocon,Sae} a small eigenvalue expansion was performed which, in combination with
group theoretic techniques, yielded the first three terms of the effective action in a perturbative expansion.
The small-eigenvalue (or, equivalently, the strong interaction) behavior of the model is accessible in this
approach. The weak interaction, large-eigenvalue regime, on the other hand, is less reliably probed in this
approach. In order to explore the full structure of the model, the exact effective action is needed.

In a previous paper, the correlation functions of this model were derived and it was demonstrated that in
the large-$N$ limit the eigenvalues still distribute according to a Wigner semicircle of rescaled radius \cite{NPT}.
Some results for the interacting theory were also derived in \cite{Tek}. The full treatment of the interacting
$\phi^4$ fuzzy theory, however, is still missing.

In the present paper we deal with this problem by analyzing the full effective action for the eigenvalues
using a self-consistent ``bootstrap" technique. Only the part of the action relevant for symmetric eigenvalue
distributions is considered, which is adequate to probe the properties of the theory in the symmetric
phase. The bootstrap technique allows for the perturbative evaluation of the action up to sixth order in the
eigenvalues, but fixes its full dependence on the quadratic invariant of the matrix (trace of its square).
For distributions close to the semicircle, this is a good approximation to the full effective action.
This effective action is shown to contain nonperturbative contributions that cannot be captured in a
perturbative treatment. We identify the large-$N$ behavior of the theory using this action and derive
the phase transition from the disordered to the symmetric ordered phase.

The organization of the paper is as follows: In section 2 we review the results of the large-$N$ planar method
and analyze the effective action in the bootstrap method. We identify the low and high eigenvalue limits of the
action and point out its nonperturbative nature. The first three terms in the perturbative expansion of the action
are in agreement with
earlier perturbative calculations. We also consider generalized kinetic terms and demonstrate the large-$N$
scaling limit of their effective action.

In section 3 we take the large-$N$ limit and, for the appropriate scaling, derive the eigenvalue distribution
for the interacting theory. The standard polynomially deformed Wigner semicircle obtains, but with a
redefined radius. For negative quadratic term, leading to a double-well potential, we identify the phase
transition from the disordered to the symmetric ordered phase, where the eigenvalue distribution splits
into two parts. We derive the critical line on the parameter plane and show that, in the weak interaction
limit, it has a nonperturbative dependence on the coupling. The results closely agree with previous numerical
investigations. Finally, we comment on the second phase transition, from the symmetric to the asymmetric
ordered or partially ordered phase.

\section{Calculation of the effective action}

\subsection{The fuzzy sphere action}

We consider the simplest case of a real scalar field on a fuzzy two-sphere.
The Cartesian coordinates are represented by $N \times N$ matrices:
\be
X_\alpha = \frac{2R}{N} L_\alpha ~, ~~ [L_\alpha , L_\beta ] = i \epsilon_{\alpha \beta \gamma} L_\gamma ~,~~
\sum_\alpha X_\alpha^2 = \left( 1 - \frac{1}{N^2} \right) R^2
\ee
where $L_\alpha$, $\alpha = 1, 2, 3$ are $SU(2)$ generators (angular momentum matrices) in the $N$-dimensional representation,
$R$ represents the radius of the sphere and $\theta = 2 R^2/N$ is the noncommutativity parameter.
Fields become general $N \times N$ matrices $M$. Derivatives (rotations) and the corresponding Laplacian are $L$-commutators
\be
{\cal L}_\alpha M = -i [L_\alpha , M] ~,~~~ \Delta M = \sum_\alpha {\cal L}_\alpha^2 M
= - \frac{1}{R^2} \sum_\alpha [L_\alpha , [L_\alpha , M]]
= - \frac{1}{R^2} C_2 (M)
\ee
where $C_2 (M)$ is the quadratic Casimir of the adjoint action of the $SU(2)$ generators $L_\alpha$ on the matrix $M$.
Integration over the sphere becomes a matrix trace
\be
\int_{S^2} d^2 x \, \Phi = \frac{4\pi R^2}{N} \Tr M
\ee
A real scalar field on the two-sphere is represented by a hermitian $N \times N$ matrix $M$.

We shall consider the case of a massive field with a quartic interaction term. The commutative Euclidean action would be
\be
S = \int d\mu (S^2)\, \left[ {1 \over 2}  (\nabla \phi )^2 + {\mu^2 \over 2} \phi^2 + {\lambda \over 4} \phi^4 \right]
\label{2}
\ee
The corresponding noncommutative Euclidean action is, then, given by
\bear
S &=& - {1\over 2} \Tr ( [L_\alpha, M]\, [L_\alpha, M] ) + {r \over 2} \Tr (M^2) + {g \over 4} \Tr (M^4 )\nonumber\\
&=&  \Tr \left[ {1\over 2} M\, C_2 (M) + {r \over 2} M^2 + {g \over 4} M^4 \right]
\label{1}
\eear
The parameters $r$ and $g$ are (related to) the square of the mass $\mu^2$ and
the interaction coupling constant $\lambda$. In the above we normalized the kinetic term coefficient to 1.
The proper $R$-dependent scaling factors, as well as different conventions for the action, 
such as a different normalization of the
individual terms or the introduction of a temperature parameter $\beta$, can be accommodated by rescaling $M$ and
redefining the parameters $r$ and $g$.

The commutative limit can be obtained by taking the proper large-$N$ limit of (\ref{1}). Classically, 
for `smooth' configurations, $M$ is replaced by a real scalar field 
$\phi$ on the two-sphere. The adjoint action of $L_\alpha$ on $M$, i.e. $[L_\alpha , M]$, becomes the gradient 
of $\phi$ and the action (\ref{1}), upon appropriate scaling, becomes the 
$\phi^4$  action on the two-sphere (\ref{2}).
Quantum mechanically, however, expectation values of observables depend on the noncommutativity parameter,
which acts as a regulator of infinities in the continuum, so the large-$N$ limit remains nontrivial even for the free
($g=0$) theory.

\subsection{Large-$N$ expectation values}

The kinetic term in the action (\ref{2}) can be diagonalized by expanding the matrix $M$ into components transforming
irreducibly under the adjoint action of the $L_\alpha$ (see \cite{NPT} for details). An $N$-dimensional matrix $M$
decomposes into a direct sum of representations of integer spin $\ell$, $\ell = 0,1,\dots,N-1$, of dimension $2\ell+1$
each:
\be
M = \sum_{\ell=0}^{N-1} \sum_{m=-\ell}^{\ell} c_{\ell,m} T_{\ell,m}
\label{Clm}
\ee
where $c_{\ell,m}$ are the coefficients of the expansion and $T_{\ell,m}$ the properly normalized matrices of the
$\ell$-adjoint representation.  The quadratic Casimir for each of them is $C_2 = \ell (\ell +1)$ and the kinetic
term becomes
\be
{1 \over 2} \sum_{\ell,m} \ell (\ell+1) ~c_{\ell,m}^2
\ee
The quadratic mass term can similarly be expressed in terms of the
$c_{\ell,m}$ and, in the absence of the quartic
interaction, we have a free theory in the variables $c_{\ell,m}$.

The kinetic term can further be generalized to arbitrary functions of the Laplacian ${\rm f}(\nabla^2 )$. This
translates to a kinetic term of the form $M {\rm f}(C_2 ) M$ with a corresponding function of the Casimir appearing. We shall
consider such generalized kinetic terms in the sequel, generically written as
\be
{1 \over 2} \Tr \left[ M K (M) \right] = {1 \over 2} \sum_{\ell,m} K(\ell) ~c_{\ell,m}^2
\label{Kell}\ee
with $K(\ell) = {\rm f}\bigl( \ell(\ell+1) \bigr)$.

Expectation values of scalar observables of the form
\be
T_{m,n} = \Tr ( M^m K(M)^n )
\ee
can be calculated in the large-$N$ limit of the free theory by taking advantage of the 
``planarity" of Wick contractions of the $c_{\ell,m}$
in that limit. We refer to \cite{NPT} for the details of the calculation, which we will not need here. The important
result is that the distribution of the eigenvalues of $M$ in the large-$N$ limit, obtained through the calculation
of $T_m = \Tr (M^m )$, is a Wigner semicircle of radius
\be
R_{\rm w} = 2 a = 2 \sqrt{{f(r) \over N}}
\ee
where the quantity $f(r)$ is defined as
\be
f(r) = \sum_{\ell=0}^{N-1} {2\ell+1 \over K(\ell) + r}
\ee
The density function (distribution) of the eigenvalues $x$ of $M$ in that limit is, then,
\be
\rho(x) = \frac{N}{2\pi a^2} \sqrt{ 4 a^2 - x^2}
\ee
with $a = \sqrt{f/N}$ as above. We also note that, in the large-$N$ limit, the summation in the definition
of $f(r)$ can be turned into an integral:
\be
f(r) = \int_0^N {2\ell \over K(\ell) + r} d\ell = N^2 \int_0^1 {2s \over K(Ns) + r} ds
\ee
For the case of the standard kinetic term $K = C_2 = Ns (Ns +1) \to N^2 s^2$ we have
\be
f(r) =  N^2 \int_0^1 {2s \over N^2 s^2 + r} ds = \ln \left( 1 + {N^2 \over r} \right)
\ee

\subsection{The effective action}

The matrix $M$ can be written in terms of its eigenvalues $x_i$ and its ``angular" degrees of freedom $U$
\be
M = U \Lambda U^{-1} ~,~~~ \Lambda = {\rm diag}\{ x_i\}
\ee
with $U$ a unitary matrix.
If we are interested only in observables that are functions of the eigenvalues of $M$, such as
$T_m = \Tr (M^m )$, we can integrate over the angular degrees of freedom $U$ to obtain
an effective action of the eigenvalues. The quadratic (mass) and quartic (interaction) terms of
the full action are invariant under unitary transformations of $M$ and thus independent of $U$.
Only the kinetic term depends on $U$. We are thus interested to calculate
\be
\int dU e^{-{1 \over 2} \Tr \left[ U \Lambda U^{-1} K (U \Lambda U^{-1} ) \right]}
= e^{-S_{eff} (\Lambda )}
\ee
Since $S_{eff}$ only depends on the set of eigenvalues of $M$, it will in general be a function of the trace
invariants of the matrix $T_n = \Tr (M^n )$. Further, since the identity matrix, corresponding to a constant
field on the fuzzy sphere, has vanishing kinetic term (we assume this true also for $K(M)$, that is, $K(\ell=0)=0$),
$S_{eff}$ depends only on the translation-invariant traces
\be
t_n = \Tr \left( M - {1 \over N} \Tr M \right)^n ~,~~~ n = 2,3,\dots
\ee
Finally, since the kinetic term is even in $M$, the effective action must obey
\be
S_{eff} ( \{ t_n \} ) = S_{eff} ( \{ (-1)^n t_n \} )
\ee 

To go further, we focus on the large-$N$ limit and note that, in this limit, the effective action is dominated
by the equilibrium configuration of the eigenvalues of $M$. Since all the other terms in the action are even
in $M$, and assuming that the vacuum also respects this condition, the eigenvalue distribution will obey
$\rho(-x) = \rho(x)$ and thus all odd moments vanish:
\be
t_{2n+1} = 0
\ee
As a result, the relevant part of the effective action in this limit is only the one depending on the
even moments, which we call $S_e$:
\be
S_e (\{ t_{2n} \} )= S_{eff} (\{ t_{2n+1} = 0 \})
\ee
The last crucial observation comes from the comparison to the results of the previous section: we know
that in the absence of interactions ($g=0$) the distribution of eigenvalues is a Wigner semicircle. This implies
that the equation of equilibrium for the eigenvalue $x_i$ as it arises from the effective action must contain
a linear term in $x_i$, arising from $S_e$, plus the usual two-body repulsive term arising from the exponentiation
of the Vandermonde measure. Any terms in the equation of motion quadratic or higher in the eigenvalue $x_i$
must come with coefficients that vanish for any semicircular eigenvalue distribution, ensuring that the Wigner
semicircle remains a solution. 
%This means that {\it only} $t_2$ can appear in $S_e$. 
This imposes constraints on the form of $S_e$. Indeed, the equation for $x_i$ would be
\be
\sum_n {\partial S_e \over \partial t_{2n}} \, 2n x_i^{2n-1} = 2 \sum_{j(\neq i)} {1 \over x_i - x_j}
\ee
Terms with $n>1$ produce cubic and higher order terms in the $x_i$ that would deform the distribution away from
Wigner's semicircle. Therefore, these terms must vanish when evaluated on a semicircular distribution.

The action $S_e$ can be expanded in a perturbation series in the eigenvalues, taking the form
\be
S_e = a_2 t_2 + ( a_4 t_4 + a_{22} t_2^2 ) + ( a_6 t_6 + a_{42} t_4 t_2 + a_{222} t_2^2 )
+ ( a_8 + a_{62} t_6 t_2 + a_{422} t_4 t_2^2 + a_{2222} t_2^4 ) + \dots
\ee
where each parenthesis represents terms contributing to a given (even) order in the eigenvalues.
For a semicircular distribution, the moments are related by
\be
t_{2n} = C_n \, t_2^n = \frac{(2n)!}{n!(n+1)!} \, t_2^n
\label{cata}
\ee
where the $C_n$ appearing above are the Catalan numbers. Imposing the conditions
\be
\left. \frac{\partial S_e}{\partial t_{2n}} \right|_w = 0 ~,~~~ n>1
\ee
where the subscript $w$ denotes that the moments are evaluated on a semicircle using (\ref{cata})
we obtain, order by order in the eigenvalues:
\be
 a_4 = 0~;~~ a_6 = a_{42} = 0 ~;~~ a_8 = a_{62} = 0 ~,~ 4 a_{44} + a_{422} = 0 ~; ~\dots
\ee
Since the coefficients of $t_2^n$ ($a_{2\dots 2}$) are, as yet, unconstrained, $S_e$ remains an
arbitrary function of $t_2$ but higher orders containing $t_4 , t_6 \dots$ are constrained. To tenth
order in the eigenvalues the action can be written as
\be
S_e = \half F( t_2 ) +( b_1 + b_2 t_2 )(t_4 - 2 t_2^2 )^2 + c (t_6 - 5 t_2^3 ) (t_4 - 2 t_2^2 ) + \dots
\ee
with $F$ an arbitrary function and $b_1 , b_2 , c$ some redefined coefficients. We see that higher
moments $t_4 , t_6$ etc. appear first at the eighth order in the eigenvalues. Therefore, the action up to
sixth order contains only $t_2$. Further, the extra terms, as well as their derivatives, vanish for
semicircular distributions for which $t_{2n} = C_n t_2^n$.

We are thus led to the conclusion that, for distributions reasonably close to the semicircle,
the effective kinetic term action $S_e$ for symmetric eigenvalue distributions can be well approximated
by a function of $t_2$ only. In what follows we shall concentrate on this part of the action. We 
denote $t_2 = t$ and define:
\be
\int dU e^{-{1 \over 2} \Tr \left[ U \Lambda U^{-1} K (U \Lambda U^{-1} ) \right]}
\simeq e^{-{1 \over 2} F(t)} ~,~~~
%t = \Tr\left( M - {1 \over N} \Tr M \right)^2 =
t = \Tr (M^2) - {1 \over N} (\Tr M )^2
\ee
where the coefficient $1/2$ was introduced to conform with standard kinetic term conventions.

\subsection{Self-consistent calculation of the effective action}

To find $F(t)$ we use a self-consistent ``bootstrap" method: we consider a theory with only the kinetic
term and a mass term and demand that the effective action give the same results as the planar calculation
for all values of the mass. Specifically, we consider
\be
S(M) = \Tr \left( {1 \over 2}  M K (M) + {1 \over 2} z M^2 \right)
\ee
where the mass squared has been denoted $z$ to stress that it is not the actual mass of the final
theory but rather an auxiliary parameter. The results of section (2.2) imply that the distribution of
eigenvalues is
\be
\rho(x) = \frac{N}{2\pi a^2} \sqrt{ 4 a^2 - x^2} ~,~~~
a = \sqrt{{f(z) \over N}}
\ee
The function $f(z)$ is defined as before in the large-$N$ limit:
\be
f(z) = \sum_{\ell=0}^{N-1} {2\ell+1 \over K(\ell) + z} =
N^2 \int_0^1 {2s \over K(Ns) + z} ds
\label{fz}
\ee

The equations of motion for the eigenvalues $x_i$ arising from the full effective action ${1 \over 2} F +
{1 \over 2} \Tr M^2$, on the other hand, are
\be
F' (t) \, x_i + z \, x_i = 2 \sum_{j(\neq i} {1 \over x_i - x_j}
\ee
In the large-$N$ limit the value of $t$ is fixed by the equilibrium distribution itself and becomes an
$x_i$-independent constant. Using standard results, the eigenvalue distribution becomes a Wigner
semicircle with radius fixed by the coefficient of the linear term:
\be
\rho(x) = {\omega \over 2\pi} \sqrt{{4N \over \omega} - x^2} ~,~~~
\omega = F' (t) + z
\ee
The value of $t$ can be calculated from the above distribution as
\be
t = \int x^2 \rho(x) dx = {N^2 \over \omega}
\label{tom}
\ee
which implies
\be
F' (t) + z = {N^2 \over t}
\label{tF}
\ee
Finally, matching the above semicircle to the one obtained with the planar calculation
we have
\be
a = \sqrt{{N \over \omega}} = \sqrt{{f(z) \over N}}
\ee
which, combined with (\ref{tom}) and (\ref{tF}) gives our final set of equations
\bear
&& F' (t) + z = {N^2 \over t} \cr
&&~ t = f(z)
\label{Ftz}
\eear
The above two equations determine the function $F(t)$ (up to an irrelevant constant)
in terms of the known function $f(z)$ through the auxiliary variable $z$. One needs to
invert the second equation to obtain $z= f^{-1} (t)$, insert in the first one and integrate.

For the case of the standard kinetic term $M K(M) = M C_2 (M)$ the function $f(z)$
was calculated as
\be
f(z) = \ln \left( 1 + {N^2 \over z} \right)
\ee
Inserting this in the second equation of (\ref{Ftz}), inverting to express $z(t)$ in terms of $t$, 
and performing the integral in the first equation we obtain
\be
F(t) = N^2 \ln {t \over 1- e^{-t}} = N^2 \left( {t \over 2} - \ln {e^{t \over 2} - e^{-{t \over 2}}
\over t} \right)
\label{Ft}
\ee
The second form of $F(t)$ serves to show that, apart from the term linear in $t$, it contains
only even powers of $t$. Its first few terms are
\be
F(t) = N^2 \left( {t \over 2} - {t^2 \over 24} + {t^4 \over 2880} + O(t^6 ) \right)
\label{texp}
\ee

It is useful to consider the asymptotic behavior of the effective potential. For small $t$ (which
corresponds to large $z$), an expansion of (\ref{fz}) in powers of $1/z$ and inversion leads
to the result
\be
F(t) \sim C\, t ~,~~~ C = \int_0^1 2 s K(Ns) ds
\label{tsmall}
\ee
This is a sort of ``Casimir energy" term, as the coefficient $C/2$ is the sum over all the zero-point
energies of the free oscillators $C_{\ell,m}$ in the decomposition (\ref{Clm}).
For $K(\ell) = \ell^2$ it reproduces the linear term in (\ref{Ft}). We also note that in the
small-$t$ limit the action remains a single trace.

The large-$t$ limit, on the other hand, is more interesting as it is {\it nonperturbative}, that is,
cannot be captured by the expansion of $F(t)$ in powers of $t$. (This is clear for the
expression (\ref{Ft}) in the case of the standard kinetic term.) Generically, this happens for
$z \to 0$. The first equation in (\ref{Ftz}) then gives the universal behavior
\be
F(t) \sim N^2 \ln t
\ee
An interesting phenomenon appears when the integral in (\ref{fz}) for $z=0$ is finite, in which
case $t(z=0)$ attains a finite value:
\be
t_{\rm max} = N^2 \int_0^1 {2s \over K(Ns)} ds
\ee
This corresponds to a sort of ``Bose condensation" (with $z$ playing the role of chemical
potential): $t$ in fact {\it does} diverge for $z=0$ due to the divergence of the zero mode
$\ell =0$ which becomes free in that limit but is not correctly taken into account by the
continuous integral, exactly as in the standard Bose condensation phenomenon. The function
$z(t)$, then, acquires the form
\be
z(t) = \left\{
\begin{tabular}{l l}
$z_f (t)$  &$t<t_{\rm max}$ \cr
0 & $t \ge t_{\rm max}$
\end{tabular} 
\right.
\ee
where $z_f (t)$ is the form of $z(t)$ that we get by inverting (\ref{fz}) for $t < t_{\rm max}$.
This seemingly creates a ``phase transition" in the form of $F(t)$, giving $F(t) = N^2 \ln t$ for all $t>t_{\rm max}$.
For such $F(t)$ the full effective action, including the Vandermonde logarithmic term,
becomes scale-invariant  and the equilibrium distribution can have any radius,
thus accommodating any value of $t$.

What in fact happens is that the effective action can be reliably
calculated in the bootstrap method only as long as $F' (t) > N^2 /t$; otherwise, the equilibrium configuration
in the presence of $F(t)$ alone does not exist. The condensation of the mode $\ell =0$ is not relevant,
as this is a nonplanar effect and does not contribute to a Wigner distribution but, rather, to a Gaussian
distribution for the zero mode.The effective action for the full range of $t$ would be obtained by allowing
the variable $z$ to vary over the complex plane in (\ref{fz}).

Since the effective action for small $t$ starts with a positive slope $C$ as in (\ref{tsmall}), it would seem that
the system is stable even for a range of negative values for the quadratic term $\mu^2 > -C$, which is in
contradiction to the instability of the model due to the instability of modes with $\ell^2 < -\mu^2$. What in fact
happens is that the Vandermonde term repels the eigenvalues and drives $t$ to larger values and over the point
where $F' (t) + \mu^2 <0$, thus destabilizing it as expected.

\subsection{Comparison with perturbative results}

It is useful to compare our results with the perturbative results obtained by O' Connor and S\"amann \cite{ocon}
as corrected in \cite{Sae}. Using a large-temperature expansion (which amounts to a small-eigenvalue
expansion) and group theoretic techniques, they obtained the effective action to third order in temperature
parameter $\beta$, which corresponds to sixth order in the eigenvalues. Since our calculated effective
action is exact up to this order (since higher moments $t_4$ and above would appear at eighth order in the
eigenvalues), we expect full agreement up to that order.

In order to compare our results with those in \cite{Sae} we need to account for different normalizations and
conventions. Specifically, the matrix model action was defined as
\be
S = \Tr \left[ 2 \beta M_s C_2 (M_s ) + \beta r_s M_s^2 + \beta g_s M_s^4 \right]
\ee
where the subscript $s$ denotes fields and parameters in the conventions of \cite{Sae}. To map to our action
we need to have the correspondence
\be
M = \sqrt{4\beta} M_s ~,~~r = {r_s \over 2} ~,~~ g = {g_s \over 4\beta}
\label{scorr}
\ee
Further, the variables
\be
c_n = {1 \over N} \Tr (M_s^n )
\ee
were used. Therefore we get the mapping
\be
t_n = \Tr (M^n ) = (4\beta)^{n \over 2} \Tr (M_s^n ) = (4\beta)^{n \over 2} N c_2
\ee
In particular, $t = 4\beta N c_2$.
Finally, the results were expressed in terms of the maximal angular momentum $\ell =N-1 \simeq N$ 
instead of $N$. Overall, the large-$N$ effective kinetic action for even distributions is
\be
S_e = {1 \over 2} F(4\beta \ell c_2 ) =  \beta \ell^3 c_2 -{\ell^2 \over 2}
\ln {e^{2\beta \ell c_2} - e^{-2\beta \ell c_2} \over 4\beta \ell c_2}
\ee
and from the power series expansion (\ref{texp}) we obtain
\be
S_e = \beta \ell^3 c_2 - \beta^2 {\ell^4 \over 3} c_2^2 + O(c_2^4 )
\ee
in full agreement with the cubic order result of \cite{Sae} upon putting all odd moments to zero ($c_1 = c_3 = \dots = 0$).

\subsection{Other forms of the kinetic term}

The method developed here can be used to calculate the effective action for other forms of the kinetic
term, as long as it is only a function of the Laplacian as in (\ref{Kell}). For instance, an action corresponding
to a ``relativistic" field with linear dispersion would correspond to $K(\ell) = \ell$.

We can consider kinetic terms that generically scale as $K(\ell ) = \ell^{\alpha}$ for arbitrary (positive)
$\alpha$. The function $f(z)$ becomes
\be
f(z) = N^2 \int_0^1 {2s \over N^{\alpha} s^{\alpha} + z} ds
= N^{2-\alpha} \int_0^1 {ds \over s^{\alpha /2} + {z N^{-\alpha}}}
\label{genf}
\ee
The form of $f(z)$ can be explicitly calculated, at least for integer $\alpha$. The inversion
$z = f^{-1} (t)$, however, cannot in general be performed analytically. Still, a perturbative
expansion of the effective action $F(t)$ as well as its hight-$t$ behavior can be derived.
We postpone a fuller treatment, along with the case of fuzzy $CP^n$, for an upcoming
publication. Here we only demonstrate the large-$N$ scaling properties of the effective action.

From (\ref{genf}) we see that
\be
t = N^{2-\alpha} {\bar f} (z N^{-\alpha} )
\ee
with ${\bar f}$ the function defined in (\ref{genf}) for $N=1$. This inverts to
\be
z = N^\alpha g( t N^{\alpha -2})
\ee
with $g = {\bar f}^{-1}$. Inserting in (\ref{Ftz}) we get
\be
F' (t) = {N^2 \over t} - N^\alpha g( t N^{\alpha -2})
\ee
and integrating we finally obtain
\be
F (t) = N^2 {\bar F} (t N^{\alpha -2}) ~,~~~ {\bar F} (s) = \int \left({1\over s} - g(s) \right) ds
\label{Fscale}
\ee
We see that $F(t)$ will scale as $N^2$, provided that $t$ scales as $N^{2-\alpha}$.

\section{Scaling and large-$N$ phase transitions}
\subsection{Effective action and scaling}

We now turn to the full theory, including the potential terms, and examine its scaling and phase transition
properties in the large-$N$ limit. The symmetric action for the eigenvalues after integrating out the 
angular degrees of freedom (and omitting terms depending on higher moments) is
\be
S = {1 \over 2} F\Bigl(\sum_i x_i^2 \Bigr) + {1 \over 2} r \sum_i x_i^2 + {1 \over 4} g \sum_i x_i^4 
- 2 \sum_{i<j} \ln | x_i - x_j |
\ee
%with $F$ as given in (\ref{Ft}) for the standard kinetic action. 
This leads to the equation of motion
\be
F' \Bigl(\sum_i x_i^2 \Bigr) \, x_i + r\, x_i + g \, x_i^3 = 2 \sum_{j(\neq i)} {1 \over x_i - x_j}
\ee
Expressing this in terms of the eigenvalue density $\rho (x)$ in the large-$N$ limit we obtain
\be
\left[ F' (t) + r \right]  x + g \, x^3 = 2 \Xint -  dy \, {\rho (y) \over x-y} ~~~~~{\rm (for}~ \rho(x) > 0{\rm )}
\label{rom}
\ee
subject to the normalization and self-consistency conditions
\be
\int \rho(x) \, dx = N ~,~~~ t = \int x^2 \rho(x) \, dx 
\label{const}
\ee

The scaling properties are determined by
requiring that all terms in the action have the same scaling in the large-$N$ limit and thus all remain relevant.
The logarithmic term scales as $N^2$. From (\ref{Fscale}) we see that $F(t)$ also scales as $N^2$
provided $t$ is of order $N^{2-\alpha}$, which fixes the scaling of $x_i \sim N^{(1-\alpha)/2}$
and thus the scaling of $r \sim N^\alpha$ and $g \sim N^{2\alpha -1}$. This fully determines the scaling
of both the matrix $M$ and the parameters $r,g$ in the large-$N$ limit.

We now focus on the standard (Laplacian) kinetic action, corresponding to $\alpha =2$, with
$F$ as given in (\ref{Ft}). Based on the scaling found above, we define
\be
{\tilde x} = N^{1/2} x ~,~~ {\tilde \rho} ({\tilde x}) = N^{-{3/2}} \rho (x) ~,~~
{\tilde F} (t) = N^{-2} F(t) ~,~~ {\tilde r} = N^{-2} r ~,~~ {\tilde g} = N^{-3} g
\ee
In terms of the rescaled tilded quantities the equation of motion (\ref{rom}), constraints (\ref{const})
and defining relation of $F$ (\ref{Ft}) remain the same but with all explicit factors of $N$ disappearing.
We shall drop tildes from now on and simply take $N=1$ in all formulae.

\subsection{Eigenvalue distribution}

The solution of equation (\ref{rom}) is well known. The resolvent
\be
u(z) = \int {\rho(x) \over z-x} dx
\ee
is analytic on the complex plane with a cut along the real axis on the support of $\rho (x)$.
Its behavior near the real axis is fixed by the equation of motion and yields $\rho(x)$:
\bear
u(x+i\epsilon) + u(x-i\epsilon) &=& \left[ F' (t) + r \right]  x + g \, x^3 \cr
u(x+i\epsilon) - u(x-i\epsilon) &=& -2\pi i \rho(x)
\eear
while its asymptotic behavior at $z \to \infty$ is fixed by the normalization and self-consistency
constraints:
\be
u(z) = {1 \over z} + {t \over z^3} + O(z^{-4})
\label{asym}
\ee
In the above, we used the fact that $\rho (x)$ is symmetric to put the coefficient of the $z^{-2}$
term, which is the first moment of $\rho$, to zero.

For $\rho (x)$ with support a single domain around zero (the ``single cut" solution), corresponding
to a contiguous distribution of eigenvalues, the solution for $u(z)$ is
\be
u(z) = {1 \over 4\pi i} \sqrt{4a^2 - z^2} \oint {\left[ F' (t) + r \right]  s + g \, s^3 \over (s-z)
\sqrt{4a^2 - s^2}} \, ds
\ee
where the contour for $s$ runs clockwise around the domain of $\rho$ on the real axis but does not
include the pole at $s=z$. The size of the distribution $a$ and the parameter $t$ are fixed by the
constraints on $\rho$ (\ref{const}) or, equivalently, by the asymptotic conditions (\ref{asym}).
The $z^{-1}$ term (normalization condition) yields the relation
\be
a^2 \left[ F' (t) + r \right] + 3 g a^4 = 1
\label{aF}
\ee
while the $z^{-3}$ term (self-consistency condition) yields $t$ as
\be
t = a^2 + g a^6
\label{ta}
\ee
The above two conditions, together with
\be
F' (t) = {1 \over t} - {1 \over e^t -1}
\ee
determine $t$ and $a$. The system is transcendental, so its solution cannot be written
in terms of elementary functions of $r$ and $g$, but it can be evaluated numerically.
Keeping $a$ as a parameter in the solution, the form of $\rho$ obtains as
\be
\rho(x) = {\sqrt{4a^2 - x^2} \over 2\pi a^2} \left( ga^2 x^2 -ga^4 +1 \right)
\label{solr}
\ee

\subsection{Phase transition}

When the quadratic term in the action is negative, $r<0$, the potential has two symmetric
minima that tend to spread the eigenvalue distribution away from zero. For some critical value
of $r$ the eigenvalue distribution separates into two disjoint distributions, signaling a third-order
phase transition. Beyond that critical value the expression (\ref{solr}) for the density is not
valid and a two-cut solution has to be obtained.

In spite of the transcendentality of the equations (\ref{aF},\ref{ta}) for $t$ and $a$, the
critical point can be calculated analytically. It will happen when
\be
\rho(x) = 0 ~~~~ {\rm or} ~~~~ g a^4 = 1
\ee
which gives $a = g^{-1/4}$. (\ref{ta}), then, gives $t = 2 g^{-1/2}$. Plugging these values
in (\ref{aF}) and solving for $r$ we eventually obtain the critical value
\be
r = -{5 \over 2} \sqrt{g} + {1 \over e^{2 \over \sqrt{g}} -1} =
 -{5 \over 2} \sqrt{g} -{1 \over 2} + {1 \over 2} \coth{1 \over \sqrt{g}}
\label{rcrit}
\ee
We observe that the above is nonperturbative in $g$, while it has a perturbative expansion
in $1/\sqrt{g}$. The second expression shows that, apart from the first two terms, it contains
only odd powers of $1/\sqrt{g}$. The first few terms in the expansion are
\be
r = -2 \sqrt{g} -{1 \over 2} + {1 \over 6 \sqrt{g}} - {1 \over 90 \sqrt{g}^3} +
O\left( {1\over \sqrt{g}^5} \right)
\label{rexp}
\ee

To compare with the perturbative results of \cite{Sae}, we note that the high-temperature
low-eigenvalue expansion of \cite{Sae} is essentially a high-$g$ expansion and thus corresponds
to the expansion (\ref{rexp}). To compare, we use the correspondence (\ref{scorr}) to obtain
\be
r_s = -{5 \over 2} \sqrt{g_s \over \beta} - 1 + \coth\left(2\sqrt{\beta \over g_s}\right)
\ee
and the perturbative expansion
\be
r_s = -2 \sqrt{g_s \over \beta} - 1 + {2 \over 3} \sqrt{\beta \over g_s} +
 O\left(\sqrt{\beta \over g_s}^3 \right)
\label{spert}
\ee
Again, we obtain agreement with \cite{Sae} up to the given order of perturbation.

Although the critical line in the $(r,g)$ plane has the same high-$g$ behavior, its behavior for small
$g$ is drastically different. The perturbative critical $r$ becomes zero at some finite value of $g$
and eventually goes to infinity, which is obviously unreasonable. Our result, however, remains
negative and goes to zero at $g=0$ in a nonperturbative way. This line matches the numerically
obtained results of \cite{numer} for the full range of values of $g$, providing an indication that
the effect of the omitted higher-moment terms in the effective action remains small even for distributions
quite far from a semicircle, such as the splitting double-well distribution at the phase transition.

We conclude by mentioning that a second phase transition happens when the symmetric 
two-cut solution becomes unstable against
an asymmetric one-cut solution where all the eigenvalues lump near one of the minima of the
potential. Identifying this phase transition would entail calculating the free energy of the symmetric
two-cut solution and comparing with the free energy of the asymmetric single-cut solution. When the
energy of the asymmetric solution falls below that of the symmetric one the phase transition occurs.

In fact, the phase transition could happen in a different way: the lowest free energy configuration
could be one in which the two wells of the potential, around each minimum, are both partially filled, with
different fractions of the eigenvalues in each. (The transition to a single-cut solution would correspond to
all the eigenvalues lumped around one of the minima.) This is a more generic, and therefore more likely
pattern of symmetry breaking and phase transition than the one to a single-cut configuration.

To identify which phase transition pattern prevails
we would need to know the free energy as a function of the fraction of eigenvalues in each
well. The calculation of asymmetric distributions, however, would require the full effective action for the kinetic term,
not just its symmetric part. Although an estimate can be made by using only the symmetric part of the
effective action, as was done in \cite{Sae}, this estimate is not so reliable as we have no control
over the effect of the non-symmetric terms of the action. The part of the action depending on higher
moments could become significant as well. A full treatment of the second phase transition
is postponed until a better handle of the problem is available.

\vskip 0.3cm
\noindent
{\underline {\bf Acknowledgments:}} I would like to thank Harold Steinacker, Christian S\"amann
and the anonymous referee for pointing out that the absence of higher moment terms in the action
needs only hold for semicircular distributions (which was missed in an earlier version of the manuscript)
and for interesting discussions.

%%%%%%%%%%%%%%%%%%%%%%%%%%%%%%%%%%%%%%%%%%%%%%%%
%%%%%%%%%%%%%%%%%%%%%%%%%%%%%%%%%%%%%%%%%%%%%%%%
%%%%%%%%%%%%%%%%%%%%%%%%%%%%%%%%%%%%%%%%%%%%%%%%

\begin{thebibliography}{99}

\bibitem{douglas}
For a review of noncommutative field theories see:
M.R. Douglas and N.A. Nekrasov, \RMP ~{\bf 73} (2001) 977;
A.P. Balachandran, {\it Pramana} {\bf 59} (2002) 359;
R.J. Szabo, Phys. Rept. {\bf 378} (2003) 207.

\bibitem{taylor}
For a review of the matrix version of M-theory and its solutions, see,
W. Taylor IV,
\RMP ~{\bf 73} (2001) 419.

\bibitem{BerMad}
F.A. Berezin, Comm. Math. Phys {\bf 40} (1975) 153; J. Madore, Class. Quant. Gravity {\bf 9} (1992) 69.

\bibitem{wigner} E. Wigner, Ann. Math. {\bf 62}, 548 (1955);
{\it ibid.} {\bf 67}, 325 (1957).

\bibitem{string}  D.J. Gross and N. Miljkovic, \PL~{\bf 238B}, 217 (1990); E. Brezin, V. Kazakov and Al.B. Zamolodchikov, \NP~{\bf B338}, 673 (1990); P. Ginsparg and J. Zinn-Justin, \PL~{\bf 240B}, 333 (1990); G. Parisi, \PL~{\bf 238B}, 209 (1990);
D.J. Gross and I.R. Klebanov, \NP~{\bf  B344}, 475 (1990);
I.R. Klebanov, ``String Theory in Two Dimensions,'' in String Theory and Quantum
Gravity '91, eds. J. Harvey et al. (River Edge, NJ: World Scientific, 1992);
D.J. Gross and A.A. Migdal, \PRL~{\bf 64}, 717 (1990); M. Douglas and S. Shenker, \NP~{\bf B335}, 635 (1990); E. Brezin and V. Kazakov, \PL~{\bf 236B}, 144 (1990).
For a review, see P. Ginsparg and G. Moore, {\it Lectures on 2D gravity and 2D string theory},
TASI lectures (1992), arXiv:hep-th/9304011.

\bibitem{CalPol}
For a recent review see A.P. Polychronakos, J. Phys. {\bf A39} (2006) 12793.
%%CITATION = HEP-TH/0607033;%%

\bibitem{QH}
L. Susskind, arXiv:hep-th/0101029;
A.P. Polychronakos, JHEP {\bf 0106} (2001) 070 and {\bf 0107} (2001) 006.
%%CITATION = HEP-TH/0103013;%%
%%CITATION = HEP-TH/0106011;%%



\bibitem{guhr} For a review with applications to condensed matter and disordered systems see T. Guhr, A. M\"uller-Groeling and H.A. Weidenm\"uller,
Phys. Rep. {\bf 299}, 189 (1998).

\bibitem{chaos} 
M.V. Berry and M. Tabor, 
Proc. R. Soc. London, Ser. {\bf A 356}, 375 (1977);
O. Bohigas, M. J. Giannoni and C. Schmit, \PRL~
{\bf 52}, 1 (1984); for a recent review, see
P. Bourgade and J.P. Keating, Seminaire Poincar\'e XIV, 115 (2010).

\bibitem{steinYM} 
H. Steinacker, \NP~{\bf B679}, 66 (2004).

\bibitem{cutoff}
S.S. Gubser and S.L. Sondhi, Nucl. Phys. {\bf B605} (2001) 395.

\bibitem{numer}
M. Panero,  SIGMA 2 (2006) 081 and JHEP 0705 (2007) 082;
F. Garcia Flores, D. O'Connor and X. Martin, hep-lat/0601012 and IJMP {\bf A24} (2009) 3917.

\bibitem{steinE}
H. Steinacker, JHEP {\bf 0503} (2005) 075.

\bibitem{ocon}
  D.~O'Connor, C.~Saemann,
 %``Fuzzy Scalar Field Theory as a Multitrace Matrix Model,''
  JHEP {\bf 0708}, 066 (2007)
  [arXiv:0706.2493 (hep-th)].

\bibitem{Sae}
C. S\"amann, SIGMA {\bf 6} (2010) 050.

\bibitem{NPT}
V.P. Nair, A.P. Polychronakos and J. Tekel, Phys. Rev. {\bf D85} (2012) 045021.
%%CITATION = ARXIV:1109.3349;%%

\bibitem{Tek}
J. Tekel, Phys. Rev. {\bf D87} (2013) 085015.



\end{thebibliography}
\end{document}